\documentclass[10pt,a4paper]{article}
\usepackage[utf8]{inputenc}
\usepackage{amsmath}
\usepackage{amsfonts}
\usepackage{amssymb}
\usepackage{graphicx}
\usepackage{hyperref}

\author{B. Ananthanarayan and Shayan Ghosh\\
Centre for High Energy Physics,\\
Indian Institute of Science, \\Bangalore 560 012, India}

\date{December 19, 2018}

\newcommand{\bea}{\begin{eqnarray}}

\newcommand{\eea}{\end{eqnarray}}
\newcommand{\beq}{\begin{equation}}
\newcommand{\eeq}{\end{equation}}

\begin{document}

\title{Pion Interactions and the Standard Model at High Precision \footnote{Prepared for
the special issue of Physics News, Quarterly Publication of the Indian Physics Association being brought out to mark the occasion of the 125th birth anniversary of  S.N. Bose.}}
\maketitle

\bigskip

{\bf Abstract:} Pions were predicted by H. Yukawa as force carriers of the inter-nucleon forces, and were detected in 1947. Today they are known to be bound states of quarks and anti-quarks of the two lightest flavours. They satisfy Bose statistics, and are the lightest particles of the strong interaction spectrum.  Determination of the parameters of the Standard Model, including
the masses of the lightest quarks, has only recently reached high precision on the lattice. Pions are also known to be pseudo-Goldstone bosons of spontaneously broken approximate axial-vector symmetries, and a probe of their properties and interactions at high precision tests our knowledge of the strong interactions. 
Despite their long history, there are significant experimental and theoretical challenges in determining their properties at high precision. Examples include the lifetime of the neutral pion, and the status of their masses and decay widths in effective field theories. Pion-pion scattering has been studied for several decades using general methods of field theory such as dispersion relations based on analyticity, unitarity and crossing. Knowledge from these theoretical methods are used to confront high precision experimental data, and to analyze them to extract information on their scattering and phase shift parameters. This knowledge is crucial for estimating the Standard Model contributions to the anomalous magnetic moment of the muon, which is being probed at Fermilab in ongoing experiments. Other sensitive tests include the rare decay of the eta meson into three pions, which represents an isospin violating decay. The present article briefly reviews these important developments.

\newpage

At a time when one did not know much about the inter-nucleon forces, i.e., the forces between protons and neutrons inside nuclei, H. Yukawa proposed the existence of certain particles associated with their finite range, to be massive spin-$0$ `bosons' known as pions. They come in three types and weigh a little over 135 MeV$/c^2$, in units where the proton and neutron are approximately 939 MeV$/c^2$. They were subsequently discovered in 1947 in cosmic rays by C. Powell and G. Occhialini\footnote{Recent fascinating historical research shows that D. M. Bose and B. Chowdhury had been conducting research in India and may have been on the track of this discovery even earlier (cf. ``The woman who could have won a Nobel: Despite being a pioneer in the study of cosmic rays in India, Bibha Chowdhuri remains practically unknown'' by Amitabha Bhattacharya, The Telegraph, November 25, 2018).}. These very light particles constitute the lightest hadrons and are the ground state of the strong interaction spectrum. In this article we will recall some of the important highlights of pion interactions which are at the heart of our modern day understanding of elementary particle physics at low and intermediate energies. The background of the topics covered here are textbook material, and good introductions are, for example, Refs.~\cite{HoKim, Burgess}. A lot of material may also be found in the detailed review sections of The Review of Particle Properties Ref.~\cite{PDG}. Some parts of this article have either appeared, or the material reviewed, in Refs.~\cite{CS,DAE,Abbas1,Rev1,Rev2}.

Recall that in a pioneering work S.N. Bose (Fig. 1) addressed the problem of statistics in quantum mechanics and formulated the laws governing particles now known as bosons. Together with A. Einstein he proposed Bose-Einstein statistics, and systems which are governed by these, among other properties exhibit the phenomenon of condensation. The earliest known bosons were photons. Fermions, on the other hand are governed by statistics known as Fermi-Dirac statistics, and the particles that make up most of known matter are fermions, including protons, neutrons and electrons. The photon is a force carrier, and there are other force carriers associated with the weak interactions known as W and Z bosons, while those associated with the strong interactions are known as gluons, and that of gravitation are known as gravitons; they are all bosons. Pions, as the earliest known bosons besides photons, have a historicity associated with the name of S.N. Bose in whose honour this article is being written.

Today, we know that protons and neutrons, collectively known as nucleons are members of a family known as baryons, while pions and kaons are known as mesons. They are all composite particles, and while the former are made up of three quarks, the latter are made of a quark and anti-quark pair. Collectively, they are all known as hadrons, viz., particles that participate in the strong interactions. We know today that the lightest in these families are made up of the u and d type quark and anti-quarks.

The strong interactions describe particles that are made up of the fundamental constituents knows as quarks and gluons. This is a gauge theory of interactions known as Quantum Chromodynamics, which was formulated in the 1970s, see Refs.~\cite{FGL,MP,Gross:1973id,Politzer:1973fx}. The quarks come in six varieties, of which the lightest two, the u and d quarks, are the constituents of stable matter, as the others decay due to the weak interactions. Baryons and mesons are the effective low-energy degrees of freedom when quark and gluon degrees freeze out. The lightness of the pions on the hadronic scale ($m_{\pi^0}\simeq 135$ MeV, $m_{\pi^+}\simeq 139$ MeV (we have set the velocity of light, $c=1$, a common convention) is today understood in terms of a phenomenon called `spontaneous symmetry breaking' of a global symmetry. W. Heisenberg had already postulated in the early 1930's that the strong interactions appear to be invariant under the exchange of protons and neutrons, which he called an isospin symmetry, viz., that nucleons behave like the up and down components of a spin vector. Angular momentum theory then tells us that there are higher spins also. In this framework, pions lie in an isospin triplet. There are specific rules for combining the scattering of particles of definite isospin if the interaction conserves isospin. Today, we understand this to arise from the near equality of the masses of the u- and d- quarks, and the fact that the electromagnetic interaction induces only small corrections as their electric charges are different. This isospin symmetric world is a theorist's paradise. The u and d quarks have heavier cousins, the next heaviest being the s quark. The heavier counterpart of the pions, which contain the strange quark, are known as kaons. These come in four types, and along with the pions and an eight particle known as the $\eta$, form what is known as the pseudo-scalar octet. This is due to the spontaneous symmetry breaking of eight axial-vector generators of the approximate global flavour symmetry to its vector subgroup: $SU(3)_L\times SU(3)_R \to SU(3)$, with each $SU(N)$ having $N^2-1$ generators, and with $N=3$ in our case, in accordance with the Nambu-Goldstone theorem. When restricted to $N=2$, which is a much better approximate symmetry, we get isospin symmetry. Note here that this phenomenon was inspired by the investigations of superconductivity in the pioneering work of Y. Nambu. Thus, at low-energies where the quarks and gluons get bound inside hadrons, the effective low-energy theory is that of nucleons and the mesons. Due to the approximate symmetries, one can analyze the theory in terms of relations between various quantities defining the theory. This is known as the programme of effective field theories. These effective theories exploiting the symmetry properties of the strong interactions to provide a consistent framework as an expansion in powers of momenta and the quark masses has come to be known as chiral perturbation theory, and is identified with the work of J. Gasser and H. Leutwyler~\cite{GLAnn,Gasser:1984gg}, following a scheme first proposed by S. Weinberg~\cite{WeinbergPhysica}.

The properties of pions and kaons, which are bound states of the strong interactions, when tested at high precision offer a test of the Standard Model (SM). Pions, which come in three varieties $\pi^+,\, \pi^-,$ and $\pi^0$, are the lightest of all strongly interacting particles, and are bound states of quark and anti-quark pairs of u and d varieties. Pions hold the key to our understanding of
the strong interaction, which resists an analytic solution in the low-energy domain. The strong interactions at the microscopic level are described by quantum chromodynamics (QCD), which is a theory
in which the degrees of freedom are the quarks and gluons, while at macroscopic length scales one observes mesons (pions, kaons, etc.) and baryons (protons, neutrons, hyperons, etc.).

\begin{figure}
\center
\includegraphics[width=.5\columnwidth]{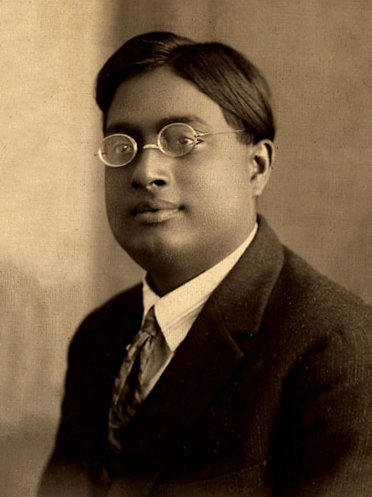}
\small{\caption{S.N. Bose, 1925}}
\end{figure}

\begin{figure}
\center
\includegraphics[width=0.9\columnwidth]{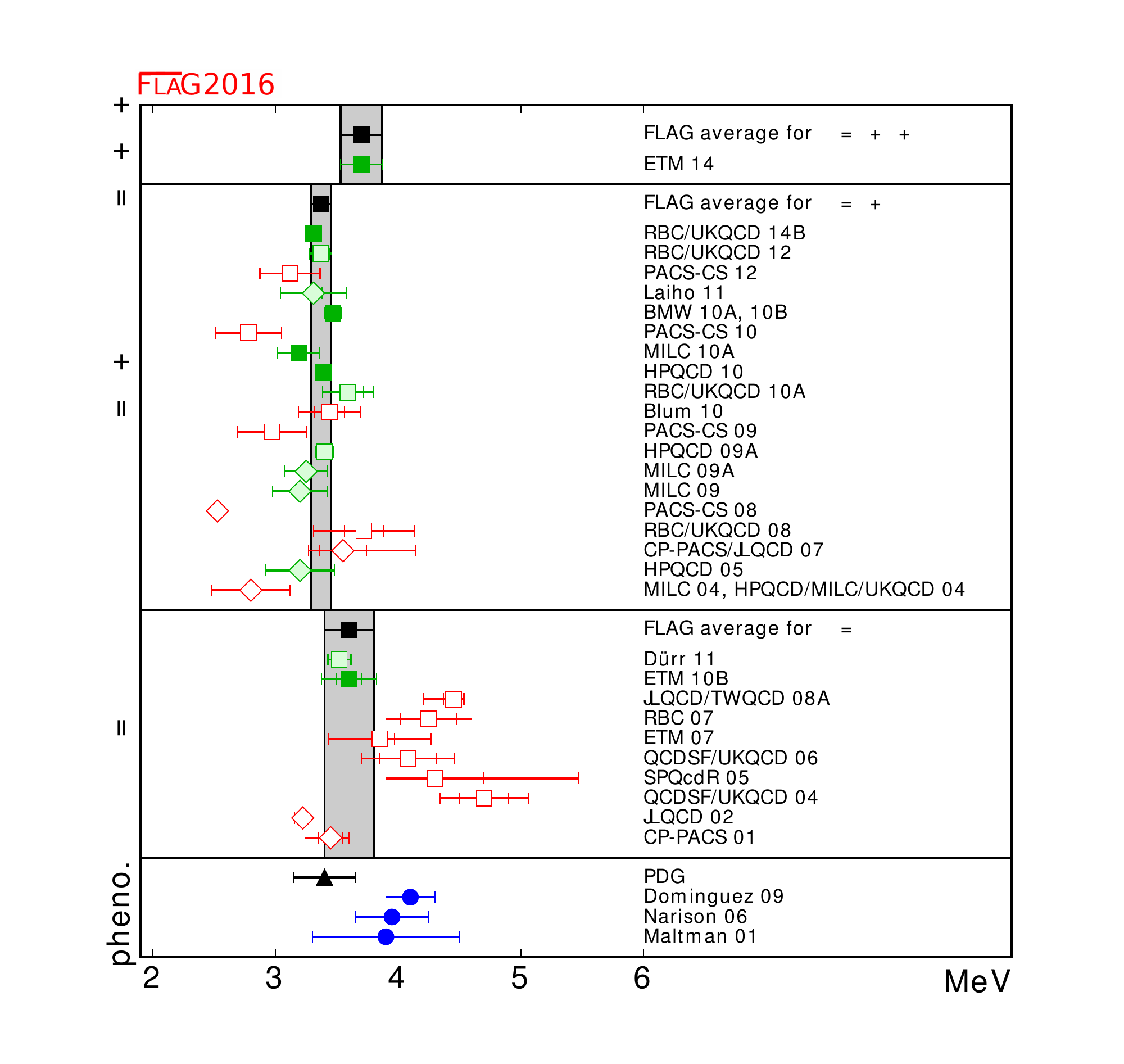}
\vspace{0.0cm}
\hspace{0.0cm}
\includegraphics[width=0.9\columnwidth]{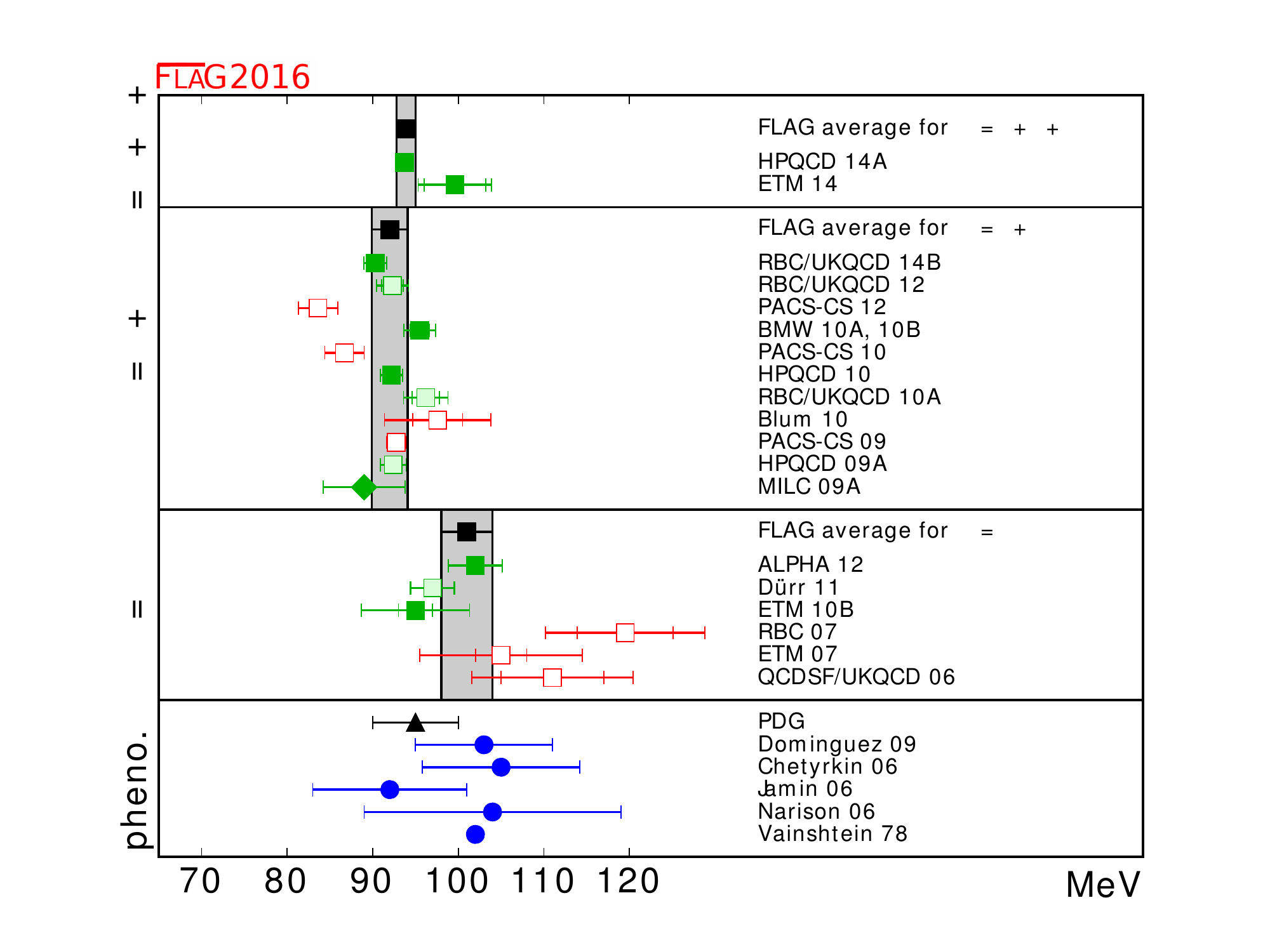}
\vspace{0.0cm}
\caption{\small{Some mass determinations from FLAG~\cite{Aoki:2016frl} of
the average lightest quark masses and the s-quark mass.  These are the
mass parameters quoted at $\mu=2$ GeV.}}
\label{fig:quarkmass}
\end{figure}

The masses of the light quarks are three of the fundamental parameters of the SM. Due to confinement, there is the issue of the definition of the quark masses itself, which has been extensively discussed in the past (for a classic work see Ref.~\cite{GL} as well as Ref.~\cite{Leutwyler:2009jg}). It has been customary to define the quark masses at the scale of 2 GeV. In the past, the best estimates came from the use of sum rules and the use of chiral perturbation theory. There has been, in the recent past, a huge effort to extract them from lattice simulations at an unprecedented level of accuracy. Several different groups using different kinds of fermions, algorithms and actions have produced numbers over a long period of time. In order to harmonize these findings and to provide the community with a coherent framework, the Flavour Lattice Averaging Group (FLAG) has now produced a series of comprehensive reports which give a detailed summary of the measurements, see Refs.~\cite{FLAG1,FLAG2,Aoki:2016frl}. The upshot of these studies is that light quarks today are somewhat lighter compared to the estimates from a couple of decades ago, as the detailed lattice simulations are most compatible with such numbers. An example from the FLAG report
is displayed in Fig.~\ref{fig:quarkmass}. Note how small the u- and d- quark masses are compared to the s quark mass.

We are now in an era when many of the properties of the strong interactions are probed at high precision on the lattice. In particular, lattice evaluations of the light quark masses have reduced error now because of advances in the field. In the past, the three lightest quark masses were evaluated using QCD sum rules and a variety of other techniques that involved phenomenological information. While the determinations are consistent with one another, the phenomenological uncertainties are significantly larger \cite{Aoki:2016frl}.  

A sensible approach requires us to match chiral perturbation theory in a controlled manner with the lattice information in three flavoured chiral perturbation theory. Pion, kaon and eta masses and decay constants were evaluated nearly two decades ago up to two-loop order \cite{Amoros:1999dp}. Nevertheless, some of the so-called sunset diagrams which are the simplest two-loop self-energy diagrams cannot all be evaluated in terms of known functions and had to be evaluated only numerically. In a series of recent publications     \cite{Ananthanarayan:2017yhz, Ananthanarayan:2018irl}, we have advanced a suitable Mellin-Barnes technique to obtain double series expansions in ratios of the masses of the three pseudoscalar mesons, which allows a controlled comparison to be performed.

One of the key processes that lies at the heart of the chiral anomaly of the light flavour sector is the neutral pion lifetime. This is fixed almost entirely by the charged pion decay constant, and the fundamental constants such as  $\alpha$ and $\hbar$. Experimental measurements of the lifetime, coming from the Primakoff process (Fig. 3) of collisions of X-rays with nuclear targets are in broad agreement with this prediction. However, isospin violation, which also manifests itself in $\pi^0-\eta$ mixing, leads to a substantial correction in the
lifetime~\cite{AM1,GBH,KM} (Fig. 4). A recent review is Ref.~\cite{BH}. The prediction for the width from the anomaly with inputs from the pion-decay constant $F_\pi$ reads:
\begin{equation}\nonumber
\Gamma(\pi^0\to 2 \gamma) = \left( \frac{M_{\pi^0}}{ 4 \pi} \right)^2 \left( \frac{\alpha}{ F_\pi} \right)^2=
7.760\, {\rm eV} \, [\tau\equiv 1/\Gamma=8.38 \times 10^{-17}\, {\rm s}].
\end{equation}
A recent high precision experiment at JLab, the Primex experiment  has measured the lifetime to the desired precision which brings theory into agreement with experiment at an unprecedented level Ref.~\cite{Larin}. This reads $\Gamma=7.82\pm 0.14({\rm stat.}) \pm 0.17({\rm syst.})$ eV, which agrees with the corrections given in Ref.~\cite{AM1,GBH,KM}, and shown in Fig.~\ref{fig:Gamma-ChPT2}. The latest standard value for this quantity, as reported in the 2018 Reviews of Particle Properties \cite{PDG} is $\tau=(8.52\pm 0.18) \times 10^{-17}\, {\rm s}$.

\begin{figure}
\center
\includegraphics[width=.7\columnwidth]{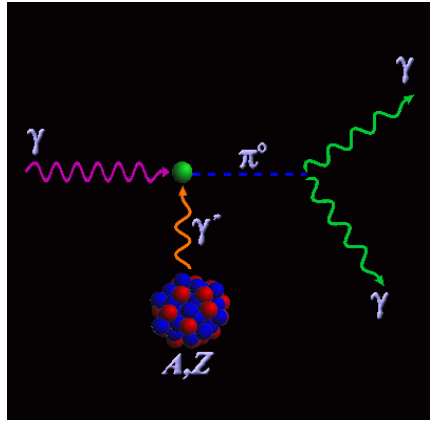}
\caption{\small{The Primakoff Effect (from JLab web-site)}}
\end{figure}

\begin{figure}
\begin{center}
\hspace{0.0cm}
\includegraphics[width=.5\columnwidth]{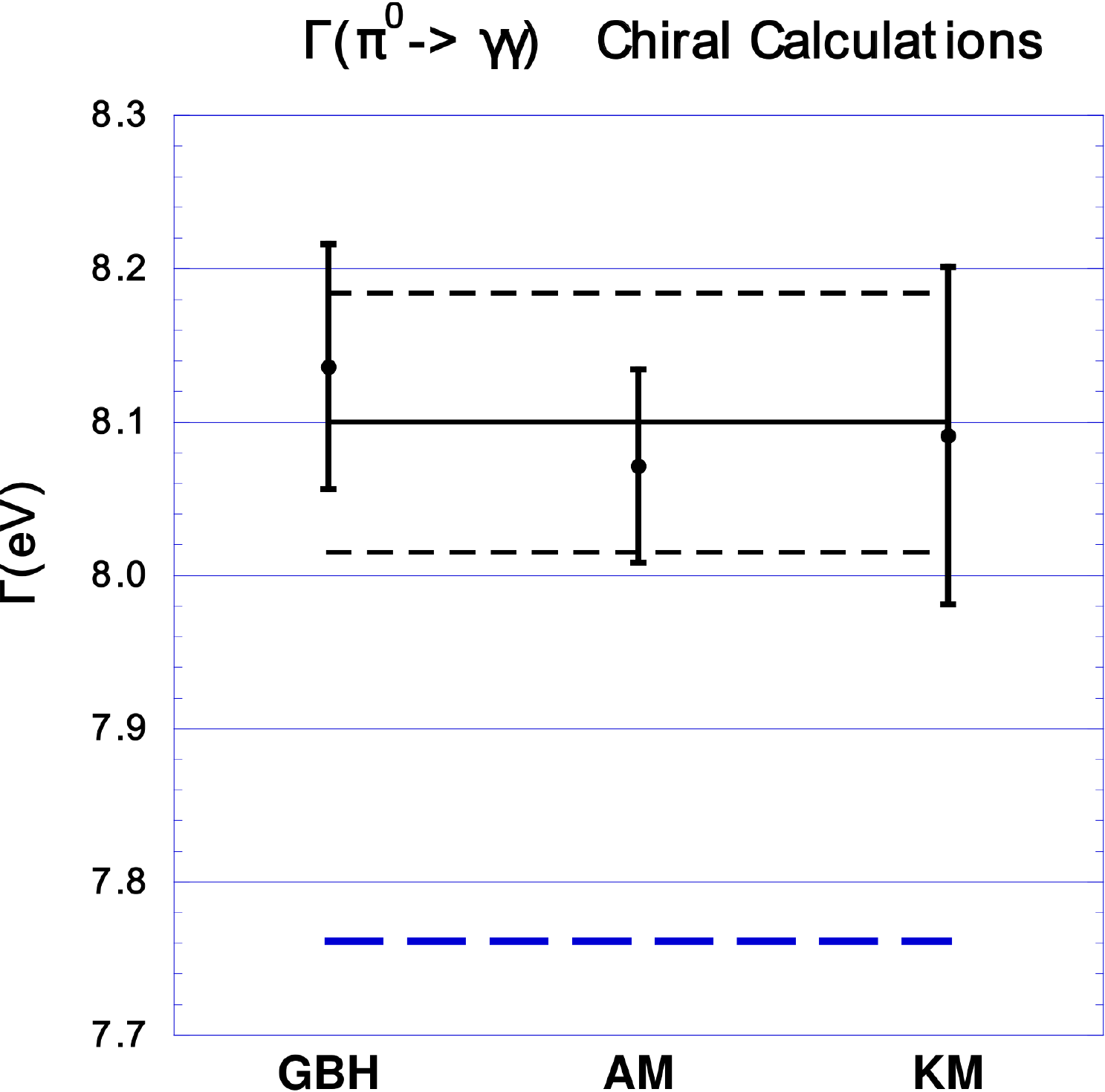}
\vspace{0.0cm}
\caption{\small{Summary of chiral corrections from Ref.~\cite{BH}. The large dashed line is the anomaly prediction, the three points shown are the predictions from Refs.~\cite{AM1,GBH,KM}, and the upper solid line is the average of these and 1\% error is the band.}}
\label{fig:Gamma-ChPT2}
\end{center}
\end{figure}

In Fig. 4,


In quantum mechanical processes involving scattering of a projectile off a target, the fundamental quantity is known as the scattering amplitude, which measures the strength of the interaction between the two particles. This is in general a function of the momentum transferred between the particles, and is a complex function of the energy involving real and imaginary parts. They also enjoy properties when viewed as functions of a complex energy variable, as well as the properties of unitarity and analyticity, which represent probability conservation and well-behavedness as a function of the energy variable, respectively. Thus, as with usual complex variables one can speak of a modulus and a phase for some of these quantities. (A closely related observable is known as the form factor, which encodes the fact that the pion is a composite particle. If it had been point-like, the form factor would simply have been unity.) These are often decomposed into partial waves, where each wave has a fixed angular momentum. Such partial waves are denoted in spectroscopic notation by S-, P-, D-, F-, G-... and higher waves. At low energies, the lowest partial waves dominate. Furthermore, near the threshold for real scattering, the real part permits an expansion in powers of the momentum, which also accounts for the centrifugal barrier, and brings in expansion coefficients known as threshold parameters. The lowest one is called a scattering length ($a^I_l,\, I=0,2$, where the subscript denoted by $I$ stands for `isospin' and the subscript $l$ denotes the fact that this is the scattering length of the angular momentum $l$ channel, $l=0$ gives the S-wave scattering length). The next threshold parameter is called the effective range. At low energies, the cross-section is simply the square of the S- wave scattering length in appropriate units. Thus, it measures the strength of the interaction. It is common in scattering experiments to decompose the scattering amplitude into `partial wave amplitudes', each of which has a definite angular momentum $l\hbar$; the S- wave corresponds to the part that is independent of the scattering angle.

Pion-pion scattering has long occupied the attention of theorists even before the advent of QCD. Of note is the programme that was launched and nurtured by A. Martin, see, e.g., Ref.~\cite{Martin}. Note that the pions lie in an `iso-triplet', corresponding to isospin 1. Therefore, pion-pion scattering amplitudes could carry isospin of 0, 1, 2 (addition of two isospins $I_1$ and $I_2$ implies that the total isospin could lie between $I_1+I_2$ and $|I_1-I_2|$). The advantage of the isospin amplitudes is that the amplitudes for all the physical processes,
$\pi^+ \pi^+ \to \pi^+ \pi^+$,
$\pi^- \pi^+ \to \pi^- \pi^+$,
$\pi^- \pi^- \to \pi^- \pi^-$,
$\pi^+ \pi^- \leftrightarrow \pi^0 \pi^0$, and
$\pi^0 \pi^0 \to \pi^0 \pi^0$ may all be expressed in terms of these, when the mass difference of the charged and neutral pions is neglected. Note that it is the property of generalized Bose statistics that requires the isospin even amplitudes to have only even angular moments (S-, D-, ... waves), and the $I=1$ to have odd angular momenta (P-, F-, ... waves).

At leading order in the low-energy expansion, S. Weinberg gave a prediction for $a^0_0$ of 0.16~\cite{Weinberg}, while at next to leading order the number was revised to $0.20\pm 0.01$, which resulted from the comprehensive analysis by Gasser and Leutwyler. The revision was found to stay stable at next to next to leading order, for a thorough discussion, see, e.g. Ref.~\cite{Colangelo:2001df}. More general scenarios would have predicted higher values of $a^0_0$~\cite{Stern:1995fw}.

Another reason for the early attention that pion-pion scattering got was that it provided a paradise for theoreticians due to the simplicity of the process, and the possibility of deploying many powerful theoretical constraints that follow from general principles. 
Furthermore, the richness of systems of hadron scattering amplitudes and their study led 
to the  rise of dual resonance theory, the Veneziano amplitude and its interpretation in terms of a boson string paving the path to the development of string theory. For a comprehensive review see Ref.~\cite{GSW1,GSW2}. Notable amongst these was the application of dispersion relations, relations which follow from the principle of causality in field theory. Loosely speaking, dispersion relations arise from the application of Cauchy's theorem of complex variable theory to scattering amplitudes, when the latter are considered as complex functions of complex energy arguments. Other principles go under the names of `crossing symmetry' and unitarity. In the context of pion-pion scattering, a system of dispersion relations were established that entailed the presence of certain unknown functions of the momentum transfer which limited the power of the dispersion relations.

In 1971, S. M. Roy (Fig. 5) of the Tata Institute of Fundamental Research used all the general properties of scattering amplitudes to eliminate all these problems, and gave a representation that required the knowledge of the two scattering lengths only, in addition to the knowledge only of the imaginary parts of the partial waves~\cite{smroy}
(see also ref.~\cite{Mahoux:1974ej}). This work has been considered a landmark contribution. This further led to a system of coupled integral equations for all the partial waves of pion-pion scattering. However, partial knowledge of the low-lying waves and some theoretical models of the higher waves could be used to produce a determination of pion scattering lengths. This program to pin down pion scattering phase shifts came to be known as Roy equation analysis. The analysis of phase shift information provided by the rare $K_{l4}$ decay and Roy equation analysis was used to pin down $a^0_0$ to the range $0.26\pm 0.05$ based on 30,000 events from the Geneval-Saclay experiment~\cite{Rosselet:1976pu}, when activity in the field stopped for a couple of decades. 
A comprehensive account of the state of the knowledge of that era is Ref.~\cite{Martin:1976mb}. After the advent of QCD and the subsequent development of low energy effective theories for pion-pion scattering, there was a resurgence of interest in the subject. We note here that a comprehensive Roy equation analysis tailored to meet the needs of modern effective field theories was 
later presented, see Ref.~\cite{ACGL}.

\begin{figure}
\center
\includegraphics[width=.5\columnwidth]{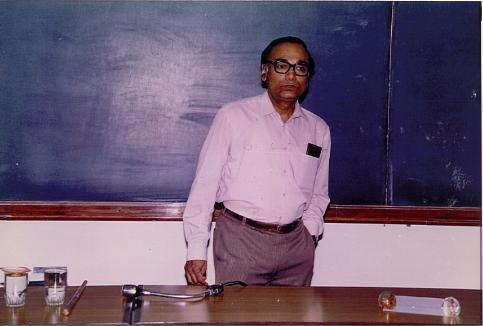}
\caption{\small{S.M. Roy (undated pictured)}}
\end{figure}

These dispersion relations are sufficiently general to permit an extension into the complex energy plane. Typically, the existence of singularities known as poles on the second Riemann sheet represents the formation of bound states of quarks and anti-quarks, and imply the formation of an intermediate unstable particle. From the real and imaginary parts of the pole position one may deduce the mass of this new particle and its `width', which is related to the inverse lifetime of the particle. Using the properties above, along with the accurately known solutions of the Roy equations, the pole position of the state known as $\sigma$ has been determined to high accuracy in the pion-pion $I=0$ channel. Note that this is a model independent way of establishing of what is the lowest-lying state in the strong interaction spectrum~\cite{Caprini:2005zr}. The mass and width are respectively given as $M_\sigma=441^{+16}_{-6}$ MeV and $\Gamma_\sigma=544^{+25}_{-18}$ MeV. Special attention may be paid to the small uncertainties. There are studies on the lattice of scattering lengths, and these have been reviewed in Ref.~\cite{hep-ph/0612112}.

We now turn to the important issue of the high precision experiments that took place in the early part of this millennium to determine scattering lengths at high precision at Brookhaven.

The NA48 collaboration in CERN has provided information through two independent measurements of combinations of so-called scattering lengths associated with the scattering of two pions. The experiment is based on in CERN's highest-intensity proton beamline, and uses a large and sophisticated detector.

The first measurement by the NA48/2 experiment~\cite{hep-ex/0511056} comes from an idea due to Cabibbo~\cite{hep-ph/0405001} to measure a `cusp' in the invariant mass distribution of pions resulting from the decay of the kaons. A cusp at an energy corresponding to $2 m_{\pi^+}$ in the number distribution of the neutral pion pair as a function of their invariant mass manifests itself as an abrupt change in the derivative of the number distribution. It is a very fine effect and can be seen only if the sample size of events that are analyzed is very large. The event sample here is enormous, and is based on about 27 million `events' $K^\pm\to \pi^\pm \pi^0 \pi^0$. In units in which the mass
of the charged pion $m_{\pi^+}$ is set to unity, they obtain for the combination of scattering lengths $|a^0_0-a^2_0|$ the value $0.264\pm 0.015$, an accurate measurement of what was a rather poorly measured experimental quantity.

The second technique employed by the NA48 collaboration comes from a rare decay of kaons in which a lepton and anti-lepton pair and two pions are produced in the decay. The re-scattering of the pions in the final state can actually be observed and provides a sensitive laboratory for the strength of the interaction of these particles. The names of the famous scientists A. Pais and S. Treiman, and those of N. Cabibbo and A. Maksymowicz are associated with this effect. Based on this technique and on the analysis of 370,000 decays a preliminary number for the scattering length $a^0_0$ is given as $0.256\pm 0.011$, according to summary talks posted on the website of the collaboration in September 2006. The E865 experiment at the Brookhaven National Laboratory, USA also uses the rare kaon decay to analyse 400,000 events and measured this quantity to be $0.216\pm 0.015$~\cite{hep-ex/0301040}. These two experiments, in addition to the low-energy phase shift, also rely on what is known as Roy equation analysis, which is described in some detail later in this article.

Another important experiment is the Di-Meson Relativistic Atom Complex (DIRAC) (Fig. 6), which uses highly sophisticated experimental techniques to get two charged pions to bind through the electromagnetic interaction to form a so-called pionium atom, that then scatters into two neutral pions in the ground state of the atom, upon which the electromagnetic interaction
is switched off and the neutral pions then scatter off. The lifetime of this state provides an accurate measurement of the same difference of two scattering lengths as the cusp experiment of NA48, an effect predicted in a different setting over 50 years ago by a highly distinguished set of authors: S. Deser, M. L. Goldberger, K. Baumann and W. E. Thirring~\cite{Deser}. Based on a harvest of 6,600 pionium atoms, the experiment reports the value of $0.264^{+0.033}_{-0.020}$~\cite{hep-ex/0504044}, from a measurement of the lifetime of the ground state of $\sim 2$ fs.  

\begin{figure}
\center
\includegraphics[width=.5\columnwidth]{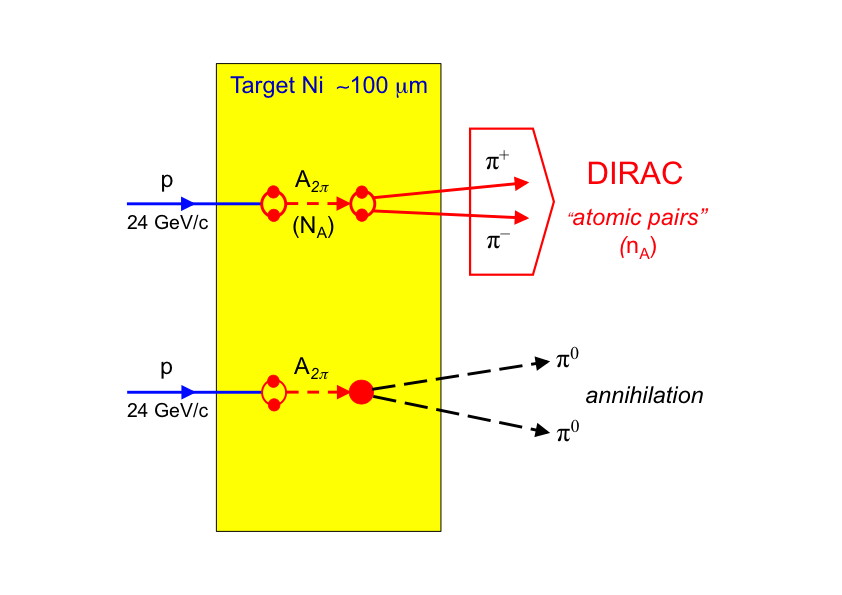}
\caption{\small{The Dirac Experiment (from CERN web-site)}}
\end{figure}

All the above said, we note here that the pion interaction offers not just a probe of the properties of the pion, but is also a crucial ingredient for evaluating the contributions of vacuum fluctuations to low energy observables, which are sensitive probes of the Standard Model, and of interactions beyond it. For instance, the pion phase shift information is crucial today for evaluating the low-energy contributions of hadrons propagating in loops to the muon $g-2$, also known as the anomalous magnetic moment of the muon, which has been measured at high precision and is again being measured at Fermilab at high precision. In this regard, one approach that has been used to improve the phenomenology here is the method of unitarity bounds and other functional methods inspired
by this approach.

It may be worth noting that field theories, effective field theories, and analysis of elementary particle physics reactions have led to the application and advancement of several branches of mathematics, including functional analysis and complex-variable techniques due to the complexity of the problems involved. Extracting strong interaction information is quite an involved subject, and some of the findings reported here are based on such modern developments. These have also had to go hand in hand with statistical methods and Monte-Carlo techniques suited to account for the statistical uncertainties of the experimental measurements. Of note is the extraction of the electromagnetic charge radius of the pion, a fundamental quantity that determines the size of pion, which is itself a composite particle. The relatively large size of the pion compared to that of the proton indicates that the motion of the quark and anti-quark is highly relativistic when bound inside the pion. The concomitant mathematical investigations have led to interesting applications in branches of mathematics such as functional analysis that are need to analyze another observable, 
the $\omega\pi$ transition form factor.
  
Form factors are important quantities of low-energy hadronic systems reflecting the fact that quarks are bound inside hadrons. Typically, these are functions of the squared momentum transfer ($t$) when quark bilinears such as electromagnetic or weak currents are sandwiched between states of the system with mesons or baryons and/or vacuum. Form factors often have known analyticity properties in the $t$ plane, with the possibility of isolated poles in some places, and well-known properties on the real axis. Their values may be known at special values of $t$ from experiment or from theoretical sources, and could obey other relations in regions of the real-$t$ axis, where one may have knowledge of the phase and/or the modulus of the form factor; or alternatively the imaginary part of the form factor, or the value of its discontinuity across the unitarity cut in the $t$ plane. There are other instances when integrals of the square of the modulus of the form factor multiplied by suitable positive-definite weight functions are either known or are bounded by using more theoretical inputs. Many of the results on form factors were obtained before the advent of QCD, and were based on very general principles; a useful review of these results is Ref.~\cite{singhraina,Abbas:2010EPJA,Rev2}. Near $t=0$, the electromagnetic form factor permits an expansion $F(t)=1 + \tfrac{1}{6} \langle r^2 \rangle t + ...$, where $\sqrt{\langle r^2\rangle}$ is called the charge radius. Of the many examples, we cite here a quantity that comes from many different sources: the
square electromagnetic charge radius has been found to have values of $\langle r^2
\rangle = 0.436(1) \text{ fm}^2$ \cite{Hanhart:2016pcd}, $0.432(4) \text{  fm}^2$ Ref.~\cite{Ananthanarayan:2017efc} and $0.429(5) \text{ fm}^2$ Ref.~\cite{Colangelo:2018mtw}.  
These considerably lower the errors quoted by the Particle Data Group~\cite{PDG}.

Experimental information on form factors comes in the form of measurements of
its modulus and phase in a variety of domains from a variety of experiments, including
electron-positron collisions, decays of the heavy $\tau$-lepton into a pion pair and
a neutrino, and indirect scattering of pions from nuclear or atomic targets from experiments at CERN, JLab, Frascati, KEK, SLAC, BEPC, Novosibirsk and CESR to name some typical examples. The property of smoothness of form factors, arising from analyticity, correlates these measurements. These are important inputs for our study. Due to lack of space we refer the reader to details of these experiments to references listed in, e.g. Ref.~\cite{Ananthanarayan:2017efc}.

Typically, the form factors lie in a class known as real-analytic functions. Common examples of this are the electromagnetic form factor of the pion, or the $\pi K$ form factor that controls the rate for the semi-leptonic decay of the kaon. In some of the examples mentioned above, the phase of the form factor is related through the Watson theorem to the phase shift of a scattering amplitude, and can be phenomenologically inferred from, e.g., the Roy equation analysis of $\pi\pi$ or $\pi K$ scattering. 
In the case of the pion electromagnetic form factor, the phase is that of the I= 1 P-wave
obtained from the solutions of the Roy equations.
This knowledge can be wired in through a suitable Lagrange multiplier technique. Lagrange multiplier techniques are also available for implementing the information on the values of the form factors where they may be real, such as at space-like points in the case of the pion form factor, or at unphysical points coming from chiral symmetry in the case of $\pi K$ scattering Ref.~\cite{Abbas:2010EPJA}. Our own work Ref.~\cite{Ananthanarayan:2017efc} uses the method of unitarity bounds and Monte-Carlo techniques to reach the high precision that is required for the charge radius. The same method has been extended in a recent work, Ref.~\cite{Ananthanarayan:2018nyx}, to determine the modulus of the form factor in four separate regimes: those probed by high precision experiments near threshold in the spacelike region, in the unphysical spacelike region which may be accessible to the lattice, to the low-energy region where experiments are not all in agreement and have large errors, and in the asymptotic spacelike regions.

Less typical are those form factors that do not lie in the real-analytic class, and an example that has recently attracted attention is the $\omega\pi$ transition form factor. In all cases, the basic relation is unitarity, which expresses the discontinuity as a product of quantities that result from the insertion of a set of states in the unitarity sum. For the pion form factor $F_\pi$, the unitarity relation in the elastic region relates the discontinuity of the form factor to the form factor itself and the P-wave of the pion-pion scattering. It follows that the phase of the form factor is equal to the phase of the pion-pion P-wave elastic amplitude, and no information about the magnitude of the discontinuity can be obtained. On the other hand, for the $\omega\pi$ form factor, the discontinuity is expressed in terms of the amplitude of the transition $\omega\pi \to \pi\pi$ and the pion form factor $F_\pi$ (to be described below). Therefore, one can calculate the discontinuity if these latter quantities are known. No information about the phase can be obtained in this case. In general, unitarity gives the discontinuity in  terms of other measurable quantities. The dispersive method uses unitarity, which allows one to express the discontinuity of the form factor in terms of the P- partial wave of the process $\pi\pi\to\omega\pi$ and the pion electromagnetic form factor. A specific feature of the $\omega\pi$  form factor, relevant for the dispersion theory, is that, due to the rescattering of the three pions that the $\omega$ can decay into, it is not a real analytic function. The treatment of rescattering effects leads to a discontinuity that does not correspond to the imaginary part of the form factor. If the small rescattering effects were to be turned off, then the problem would amount to one where the imaginary part of the form factor would be known along the elastic part of the unitarity cut, Refs.~\cite{Anant:2014,confproc}.It may be mentioned that the work above has led to an enrichment of the theory of unitarity bounds, with applications in analytic interpolation theory of functional analysis.


\begin{figure}
\center
\includegraphics[width=.5\columnwidth]{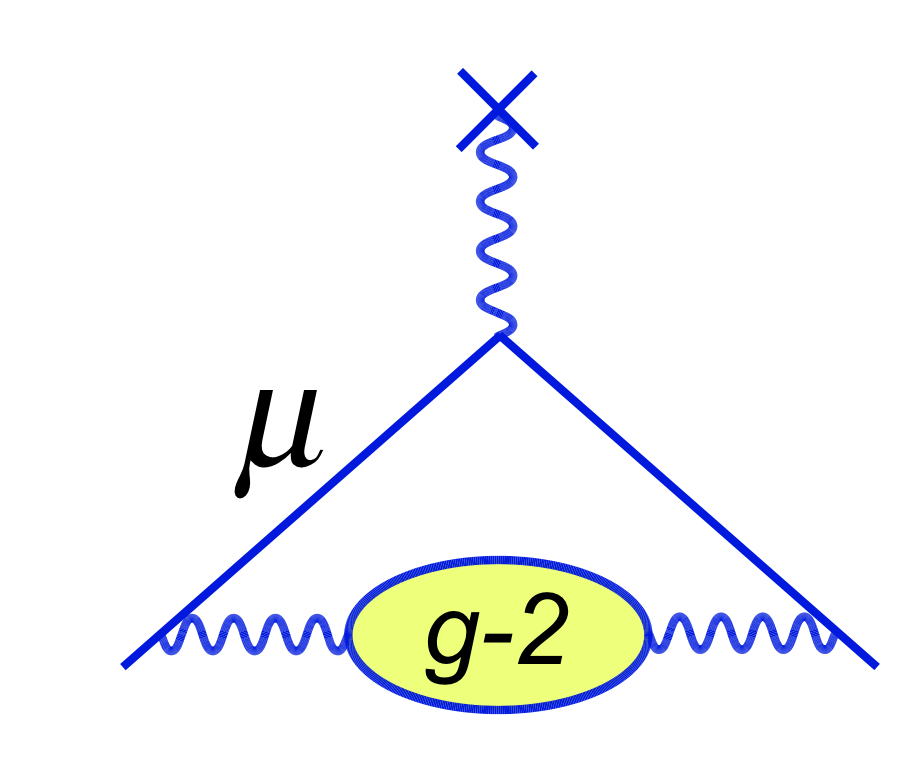}
\caption{\small{The muon $g-2$ (from BNL web-site)}}
\end{figure}

The anomalous magnetic moment (Fig. 7) is an important low energy property, see 
Refs.~\cite{Jegerlehner:2017gek,JegNyf}. One of the important bugbears in the community of the last decades has been the observed inconsistency between the experimental value of the anomalous magnetic moment of the muon from the Brookhaven experiment, which gives $a_\mu^{EXP} (\equiv {(g-2)/2})=11659208.9(6.3)\times 10^{-10}$, and its value in the Standard Model. The discrepancy is estimated to be at the level of 3$\sigma$. It may be noted that the theoretical errors are also large due to prevailing uncertainty in the hadronic contributions, which are also plagued by different data sets giving rise to different values. Improvement by a factor of four is envisaged for the future Fermilab experiment. This experiment will use a 14-m diameter electromagnet (Fig. 8) which has moved from Brookhaven to Fermilab in 2013, and which has met a new milestone with reassembly in summer 2014. The principle of the experiment is to have a highly uniform, essentially pure, 1.45 T dipole field throughout the circumference, and 3.1 GeV muons produced from pion-decay enter the magnetic field along a nearly field-free path. The $(g-2)_\mu$ measurement will be based on polarized muons being injected, and precession being studied, with parity violating weak decay being the spin analyser. Since the electrons from the decay have less energy than the muon, they curl into the storage ring and their arrival time would be measured as a function of time after injection, and oscillation on top of an exponentially falling rate gives the precession frequency, as has been recently discussed in an article in the CERN Courier from October 2014. An exciting new possibility for the measurement of $(g-2)_\mu$ is based on ultra-cold muons, see Ref.~\cite{Iinuma:2011zz}.

We note here that whereas the theoretical error is dominated by hadronic uncertainties, hadronic light by light scattering has recently been given a solid dispersive framework, see Ref.~\cite{CHKPS}. The status of the contributions of the
hadronic vacuum polarization contributions
  pion electromagnetic form factor may be found in Ref.~\cite{Benayoun}, while our analysis based on the method of functional theoretic methods which aims to lower the hadronic uncertainty may be found in Ref.~\cite{ACDI}. Our sequel to that work, which uses Monte-Carlo methods is Ref.~\cite{Ananthanarayan:2016mns}, and which gives from below 0.63~GeV the value $(133.258 \pm 0.723)\times 10^{-10}$, amounting to a reduction of the theoretical error by about $6\times 10^{-11}$.

\begin{figure}
\center
\includegraphics[width=.5\columnwidth]{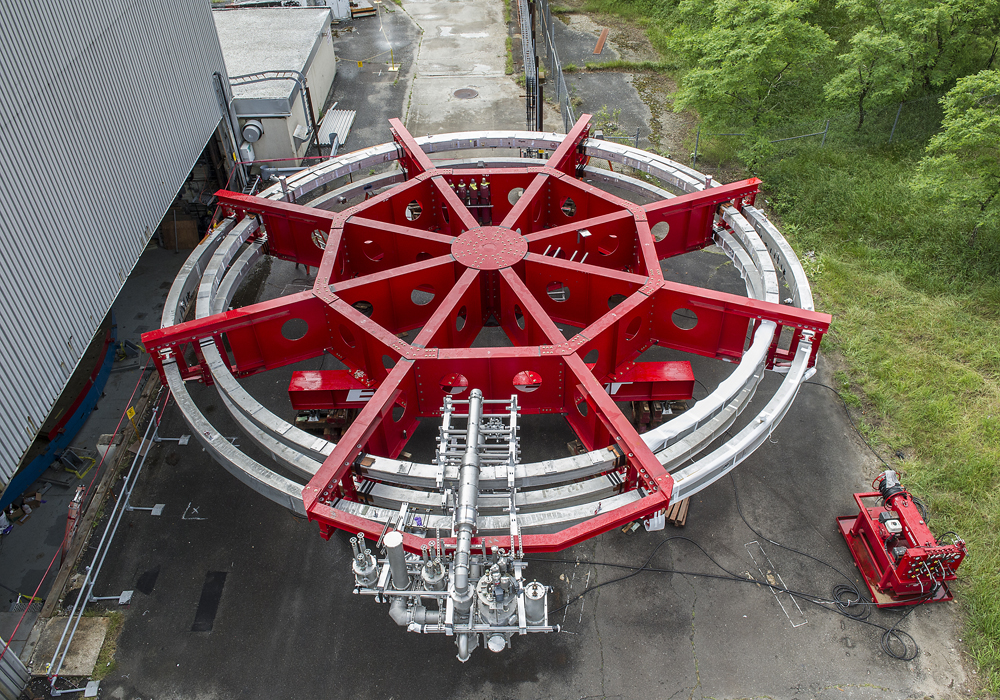}
\caption{\small{The Muon g-2 ring shortly after being removed from the building it has been in since 2001 on the Brookhaven National Laboratory site on Long Island (from BNL web-site) before being moved by land, sea and river to Illinois.}}
\end{figure}

In order to gain an understanding of isospin violation, the best sources are the well-known process $\eta\to 3 \pi$, as well as the $K^+ -K^0$ mass difference. One needs to consider the double ratio
\begin{eqnarray*}
& \displaystyle Q^2=\frac{m_s^2-\hat{m}^2}{m_d^2-m_u^2},\, \hat{m}=\frac{m_u+m_d}{ 2} & \\
\end{eqnarray*}
which gives an ellipse in the $m_u/m_d$ vs. $m_s/m_d$ plane, and can be rewritten as
\begin{eqnarray*}
& \displaystyle
\frac{m_u}{ m_d}=\frac{4 Q^2-S^2+1}{4 Q^2 + S^2 -1},\,~~ S=\frac{m_s}{\hat{m}}. & \\
\end{eqnarray*}
The most recent evaluation reads: $Q=21.3\pm 0.6$~\cite{CLLP}, which is consistent with the estimate from the violation of Dashen's theorem (the electromagnetic mass difference of kaons and pions) Ref.~\cite{AM2}. A very recent state of the art that thoroughly studies the three-body rescattering problem in this context is Ref.~\cite{Colangelo:2018jxw}, which updates the number to $Q=22.1(7)$, and provides an encyclopaedic list of references
on the subject.

As closing remarks, we point out that the pion interaction is a fundamental elementary particle physics property which controls the behaviour of the light quark sector. It plays a role in the extraction of the masses of the quarks, and has spurred effective field theories and developments in lattice gauge theory. The scattering of pions also controls the contribution of hadrons to the $g-2$ of the muon, and controls the charge radius. The pion-pion interactions also provide a window to the s quark mass. In this article we have recalled some of the recent significant probes of their property in light of experiments, as well as their impact on our own professional research. Furthermore, due to the complexity of the subject, a vast development of mathematical tools has also taken place to analyze and extract information from the experimental data. It has led to developments
in the analysis of quantities on the computer in what is known as lattice gauge theory, as well as mathematical methods based on functional analysis, and the evaluation of complicated Feynman diagrams which may or may not be amenable to analytic tools, due to which one often has to take recourse to numerical methods. There continues to be vigorous research with improved analysis of scattering information, see for example Ref.~\cite{Madrid,CCL,Elvira} for a sample of recent investigations, or extensions of the system to the related Khuri-Treiman equations, Ref.~\cite{Albaladejo:2018gif}. Pion-pion scattering has also been recently studied in the context of the bootstrap programme, see Ref.~\cite{Penedones} \footnote{The subject appears to have come full circle now from hadron physics to boot-strap to string theory and back to hadron physics and
boot-strap and string theory! A fascinating interview on the
occasion of `String Theory at 50' with G. Veneziano may be found in the 
November 2018 issue of the CERN Courier}.

\vfill

\section*{Acknowledgements}

The authors would like to thank D. Ghosh and S. Kailas, Consulting Editor and Chief Editor, respectively, of Physics News, for the invitation to write this article for the publication.
We thank S. M. Roy for proposing the topic of this article.  We also thank I. Caprini for
a careful reading and comments on the manuscript

\section*{About the authors (Fig. 9)}

B. Ananthanarayan is Professor and former Chairman of the Centre for High Energy Physics, Indian Institute of Science, Bangalore. His interests lie in several areas of elementary particle physics and field theory, including applications of low energy physics at high precision, beyond the standard model physics at colliders, as well as mathematical problems arising in 
field theory.

Shayan Ghosh recently completed his Ph.D. from the Centre for High Energy Physics, Indian Institute of Science, during which he did work furthering the Mellin-Barnes technique to analytically evaluate multi-loop multi-scale diagrams and apply them to ChPT and QED quantities.  He will pursue 
post-doctoral research at the University of Bonn in Germany.

\begin{figure}
\center
\includegraphics[width=.25 \columnwidth]{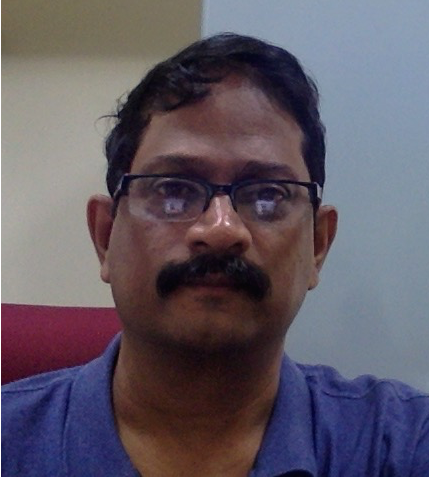}
\vspace{0.0cm}
\hspace{3.0cm}
\includegraphics[width=.25 \columnwidth]{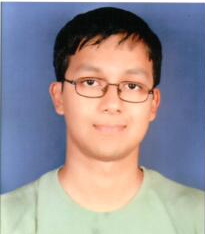}
\vspace{0.0cm}
\small{\caption{B. Ananthanaryan(left), Shayan Ghosh (right) [2018]}}
\end{figure}


\bigskip

\begin{thebibliography}{abc}

\bibitem{HoKim}
  H.~K.~Quang and X.~Y.~Pham,
  ``Elementary particles and their interactions: concepts and phenomena,''
  Berlin, Germany: Springer-Verlag, 1998. 

\bibitem{Burgess}
  C.~P.~Burgess and G.~D.~Moore,
  ``The standard model: A primer,''
Cambrdige, UK, 2011,
 
\bibitem{PDG}
  M.~Tanabashi {\it et al.} [Particle Data Group],
  Phys.\ Rev.\ D {\bf 98} (2018) no.3,  030001.


\bibitem{CS}
  B.~Ananthanarayan,
  Curr.\ Sci.\  {\bf 92} (2007) 886

\bibitem{DAE}
  B.~Ananthanarayan,
  Springer Proc.\ Phys.\  {\bf 174} (2016) 3.

\bibitem{Abbas1}
  G.~Abbas, B.~Ananthanarayan, I.~Caprini, I.~Sentitemsu Imsong and S.~Ramanan,
  Eur.\ Phys.\ J.\ A {\bf 45} (2010) 389

\bibitem{Rev1}
  B.~Ananthanarayan and I.~Caprini,
  J.\ Phys.\ Conf.\ Ser.\  {\bf 374} (2012) 012011

\bibitem{Rev2}
  B.~Ananthanarayan, I.~Caprini and B.~Kubis,
  Int.\ J.\ Mod.\ Phys.\  {\bf 31} (2016) no.14n15,  1630020.

\bibitem{FGL}
  H.~Fritzsch, M.~Gell-Mann and H.~Leutwyler,
  Phys.\ Lett.\  {\bf 47B} (1973) 365.

\bibitem{MP}
  W.~J.~Marciano and H.~Pagels,
  Phys.\ Rept.\  {\bf 36} (1978) 137.

\bibitem{Gross:1973id}
  D.~J.~Gross and F.~Wilczek,
  Phys.\ Rev.\ Lett.\  {\bf 30} (1973) 1343.

\bibitem{Politzer:1973fx}
  H.~D.~Politzer,
  Phys.\ Rev.\ Lett.\  {\bf 30} (1973) 1346.

\bibitem{GLAnn}
 J.~Gasser and H.~Leutwyler,
  Annals Phys.\  {\bf 158} (1984) 142.

\bibitem{Gasser:1984gg}
  J.~Gasser and H.~Leutwyler,
  Nucl.\ Phys.\  B {\bf 250}, 465 (1985).

\bibitem{WeinbergPhysica}
  S.~Weinberg,
  Physica A {\bf 96} (1979) no.1-2,  327.

\bibitem{GL}
  J.~Gasser and H.~Leutwyler,
  Phys.\ Rept.\  {\bf 87} (1982) 77.

\bibitem{Leutwyler:2009jg}
  H.~Leutwyler,
  PoS CD {\bf 09} (2009) 005

\bibitem{FLAG1}
  G.~Colangelo, S.~Durr, A.~Juttner, L.~Lellouch, H.~Leutwyler, V.~Lubicz, S.~Necco and C.~T.~Sachrajda {\it et al.},
  Eur.\ Phys.\ J.\ C {\bf 71} (2011) 1695

\bibitem{FLAG2}
  S.~Aoki, Y.~Aoki, C.~Bernard, T.~Blum, G.~Colangelo, M.~Della Morte, S.~Dürr and A.~X.~El Khadra {\it et al.},
  Eur.\ Phys.\ J.\ C {\bf 74} (2014) 9,  2890

\bibitem{Aoki:2016frl}
  S.~Aoki {\it et al.},
  Eur.\ Phys.\ J.\ C {\bf 77} (2017) no.2,  112

\bibitem{Amoros:1999dp}
  G.~Amoros, J.~Bijnens and P.~Talavera,
  Nucl.\ Phys.\ B {\bf 568} (2000) 319

\bibitem{Ananthanarayan:2017yhz}
  B.~Ananthanarayan, J.~Bijnens and S.~Ghosh,
  Eur.\ Phys.\ J.\ C {\bf 77} (2017) no.7,  497

\bibitem{Ananthanarayan:2018irl}
  B.~Ananthanarayan, J.~Bijnens, S.~Friot and S.~Ghosh,
  Phys.\ Rev.\ D {\bf 97} (2018) 114004

\bibitem{AM1}
  B.~Ananthanarayan and B.~Moussallam,
  JHEP {\bf 0205} (2002) 052

\bibitem{GBH}
  J.~L.~Goity, A.~M.~Bernstein and B.~R.~Holstein,
  Phys.\ Rev.\ D {\bf 66} (2002) 076014

\bibitem{KM}
  K.~Kampf and B.~Moussallam,
  PoS CD {\bf 09} (2009) 039.

\bibitem{BH}
  A.~M.~Bernstein and B.~R.~Holstein,
  Rev.\ Mod.\ Phys.\  {\bf 85} (2013) 49

\bibitem{Larin}
  I.~Larin {\it et al.}  [PrimEx Collaboration],
  Phys.\ Rev.\ Lett.\  {\bf 106} (2011) 162303

\bibitem{Martin}
  A.~Martin,
  ``Scattering Theory: Unitarity, Analyticity and Crossing,''
  Lect.\ Notes Phys.\  {\bf 3} (1969) 1.

\bibitem{Weinberg}
  S.~Weinberg,
  Phys.\ Rev.\ Lett.\  {\bf 17}, 616 (1966).

\bibitem{Colangelo:2001df}
  G.~Colangelo, J.~Gasser and H.~Leutwyler,
  Nucl.\ Phys.\  B {\bf 603}, 125 (2001)

\bibitem{Stern:1995fw}
  J.~Stern,
  arXiv:hep-ph/9510318.

\bibitem{GSW1}
  M.~B.~Green, J.~H.~Schwarz and E.~Witten,
  ``Superstring Theory. Vol. 1: Introduction,''
  Cambridge, Uk: Univ. Pr. ( 1987) 469 P. ( Cambridge Monographs On Mathematical Physics)

\bibitem{GSW2}
  M.~B.~Green, J.~H.~Schwarz and E.~Witten,
  ``Superstring Theory. Vol. 2: Loop Amplitudes, Anomalies And Phenomenology,''
  Cambridge, Uk: Univ. Pr. ( 1987) 596 P. ( Cambridge Monographs On Mathematical Physics)

\bibitem{smroy}
  S.~M.~Roy,
  Phys.\ Lett.\  B {\bf 36}, 353 (1971).
\bibitem{Mahoux:1974ej}
  G.~Mahoux, S.~M.~Roy and G.~Wanders,
  Nucl.\ Phys.\ B {\bf 70} (1974) 297.

\bibitem{Rosselet:1976pu}
  L.~Rosselet {\it et al.},
  Phys.\ Rev.\  D {\bf 15}, 574 (1977).

\bibitem{Martin:1976mb}
  B.~R.~Martin, D.~Morgan and G.~Shaw,
  ``Pion Pion Interactions in Particle Physics,''
  London 1976, 460p

\bibitem{ACGL}
  B.~Ananthanarayan, G.~Colangelo, J.~Gasser and H.~Leutwyler,
  Phys.\ Rept.\  {\bf 353}, 207 (2001)

\bibitem{Caprini:2005zr}
  I.~Caprini, G.~Colangelo and H.~Leutwyler,
  Phys.\ Rev.\ Lett.\  {\bf 96}, 132001 (2006)

\bibitem{hep-ph/0612112}
  H.~Leutwyler,
  arXiv:hep-ph/0612112.

\bibitem{hep-ex/0511056}
  J.~R.~Batley {\it et al.}  [NA48/2 Collaboration],
  Phys.\ Lett.\  B {\bf 633}, 173 (2006)

\bibitem{hep-ph/0405001}
  N.~Cabibbo,
  Phys.\ Rev.\ Lett.\  {\bf 93}, 121801 (2004)

\bibitem{hep-ex/0301040}
  S.~Pislak {\it et al.},
  Phys.\ Rev.\  D {\bf 67}, 072004 (2003)

\bibitem{Deser}
S.~Deser et al., Phys.\ Rev.\ {\bf 96}, 774 (1954).

\bibitem{hep-ex/0504044}
  B.~Adeva {\it et al.}  [DIRAC Collaboration],
  Phys.\ Lett.\  B {\bf 619}, 50 (2005)

\bibitem{singhraina}
  V.~Singh and A.~K.~Raina,
  Fortsch.\ Phys.\  {\bf 27} (1979) 561.

\bibitem{Abbas:2010EPJA}
  G.~Abbas, B.~Ananthanarayan, I.~Caprini, I.~Sentitemsu Imsong and S.~Ramanan,
  Eur.\ Phys.\ J.\ A {\bf 45} (2010) 389

\bibitem{Hanhart:2016pcd}
  C.~Hanhart, S.~Holz, B.~Kubis, A.~Kupść, A.~Wirzba and C.~W.~Xiao,
  Eur.\ Phys.\ J.\ C {\bf 77} (2017) no.2,  98
   Erratum: [Eur.\ Phys.\ J.\ C {\bf 78} (2018) no.6,  450]

\bibitem{Ananthanarayan:2017efc}
  B.~Ananthanarayan, I.~Caprini and D.~Das,
  Phys.\ Rev.\ Lett.\  {\bf 119} (2017) no.13,  132002

\bibitem{Colangelo:2018mtw}
  G.~Colangelo, M.~Hoferichter and P.~Stoffer,
  arXiv:1810.00007 [hep-ph].
  
\bibitem{Ananthanarayan:2018nyx}
  B.~Ananthanarayan, I.~Caprini and D.~Das,
 Phys.\ Rev. \ D {\bf 98} (2018) 11, 114015 
  arXiv:1810.09265 [hep-ph].
  
\bibitem{Anant:2014}
  B.~Ananthanarayan, I.~Caprini and B.~Kubis,
  Eur.\ Phys.\ J.\ C {\bf 74} (2014) 12,  3209

\bibitem{confproc}
  B.~Ananthanarayan, I.~Caprini and B.~Kubis, 
  PoS CD {\bf 15} (2016) 044.

\bibitem{Jegerlehner:2017gek}
  F.~Jegerlehner,
  ``The Anomalous Magnetic Moment of the Muon,''
  Springer Tracts Mod.\ Phys.\  {\bf 274} (2017) pp.1.

\bibitem{JegNyf} 
  F.~Jegerlehner and A.~Nyffeler,
  Phys.\ Rept.\  {\bf 477} (2009) 1


\bibitem{Iinuma:2011zz}
  H.~Iinuma [J-PARC New g-2/EDM experiment Collaboration],
  J.\ Phys.\ Conf.\ Ser.\  {\bf 295} (2011) 012032.

\bibitem{CHKPS}
  G.~Colangelo, M.~Hoferichter, B.~Kubis, M.~Procura and P.~Stoffer,
  Phys.\ Lett.\ B {\bf 738} (2014) 6

\bibitem{Benayoun}
  M.~Benayoun, J.~Bijnens, T.~Blum, I.~Caprini, G.~Colangelo, H.~Czyż, A.~Denig and C.~A.~Dominguez {\it et al.},
  arXiv:1407.4021 [hep-ph].

\bibitem{ACDI}
  B.~Ananthanarayan, I.~Caprini, D.~Das and I.~S.~Imsong,
  Phys.\ Rev.\ D {\bf 89} (2014) 3,  036007

\bibitem{Ananthanarayan:2016mns}
  B.~Ananthanarayan, I.~Caprini, D.~Das and I.~Sentitemsu Imsong,
  Phys.\ Rev.\ D {\bf 93} (2016) no.11,  116007

\bibitem{CLLP}
  G.~Colangelo, S.~Lanz, H.~Leutwyler and E.~Passemar,
  PoS EPS {\bf -HEP2011} (2011) 304.

\bibitem{AM2}
  B.~Ananthanarayan and B.~Moussallam,
  JHEP {\bf 0406} (2004) 047

\bibitem{Colangelo:2018jxw}
  G.~Colangelo, S.~Lanz, H.~Leutwyler and E.~Passemar,
  Eur.\ Phys.\ J.\ C {\bf 78} (2018) no.11,  947

\bibitem{Madrid}
  R.~Garc\'ia-Mart\'in, R.~Kami\'nski, J.~R.~Pel\'aez, J.~Ruiz de Elvira and F.~J.~Yndur\'ain,
  Phys.\ Rev.\ D {\bf 83} (2011) 074004

\bibitem{CCL}
  I.~Caprini, G.~Colangelo and H.~Leutwyler,
  Eur.\ Phys.\ J.\ C {\bf 72} (2012)  1860

\bibitem{Elvira}
  J.~Ruiz de Elvira and E.~Ruiz Arriola,
  Eur.\ Phys.\ J.\ C {\bf 78} (2018) no.11,  878

\bibitem{Albaladejo:2018gif}
  M.~Albaladejo {\it et al.} [JPAC Collaboration],
  Eur.\ Phys.\ J.\ C {\bf 78} (2018) no.7,  574

\bibitem{Penedones}
  A.~L.~Guerrieri, J.~Penedones and P.~Vieira,
  arXiv:1810.12849 [hep-th].


\end{thebibliography}
\end{document}